\documentclass{blois}

\def\prd{Phys. Rev. D}
\def\apj{Astrophys. J.}
\def\apjl{Astrophys. J. Lett.}
\def\mnras{Mon. Not. R. Astron. Soc.}
\def\prl{Phys. Rev. Lett.}
\def\jcap{J. Cosmol. Astropart. Phys.}

\def\be{\begin{equation}}
\def\ee{\end{equation}}
\def\bea{\begin{eqnarray}}
\def\eea{\end{eqnarray}}

\def\dd{\mathrm{d}} 

\newcommand{\Photo}{\includegraphics[height=35mm]{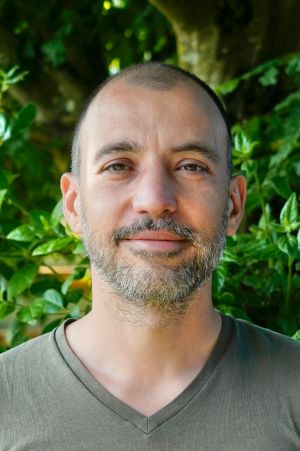}}

\begin{document}
\vspace*{4cm}
\title{What do we know about cosmic rays with energies above 5\,EeV?}

\author{Jonathan Biteau,$^{1,\,2}$ for the Pierre Auger Collaboration$^3$}

\address{$^{1}$\,Universit\'e Paris-Saclay, CNRS/IN2P3, IJCLab, 91405 Orsay, France\\
$^{2}$\,Institut Universitaire de France (IUF), France\\
$^3$ A full list of authors and affiliations can be found at \url{https://www.auger.org/archive/authors_2024_11.html}}

\maketitle\abstracts{
Cosmic rays begin to reveal their secrets at energies above 5\,EeV. Beyond this characteristic energy, known as the spectral ``ankle'', the arrival-direction data from the Pierre Auger Observatory show anisotropy on large angular scales of increasing amplitude with energy. This discovery provides observational evidence that cosmic rays beyond the ankle originate outside the Milky Way, as expected from the weak Galactic confinement and the high luminosity required for the sources. Synthetic models of extragalactic source populations emitting fully ionized atoms have allowed us to reproduce the cosmic-ray flux beyond the ankle for almost a decade. These models capture the various slope breaks in the spectrum at ultra-high energies, including the flux suppression at ${\sim}\,$45\,EeV and the recently measured feature at ${\sim}\,$15\,EeV, known as the spectral ``instep''. Such slope breaks are understood as changes in nuclear composition, with the average atomic mass increasing with energy. The population of astrophysical sources responsible for accelerating these nuclei remains unidentified, although serious contenders have been identified. Particularly instructive are the latest searches at the highest energies for anisotropies correlated with the flux patterns expected from galaxies outside the Local Group, which are approaching $5\,\sigma$.}

\section{Extragalactic photon, neutrino and cosmic-ray backgrounds}

\begin{figure}[t!]
\centering
\includegraphics[width=\linewidth]{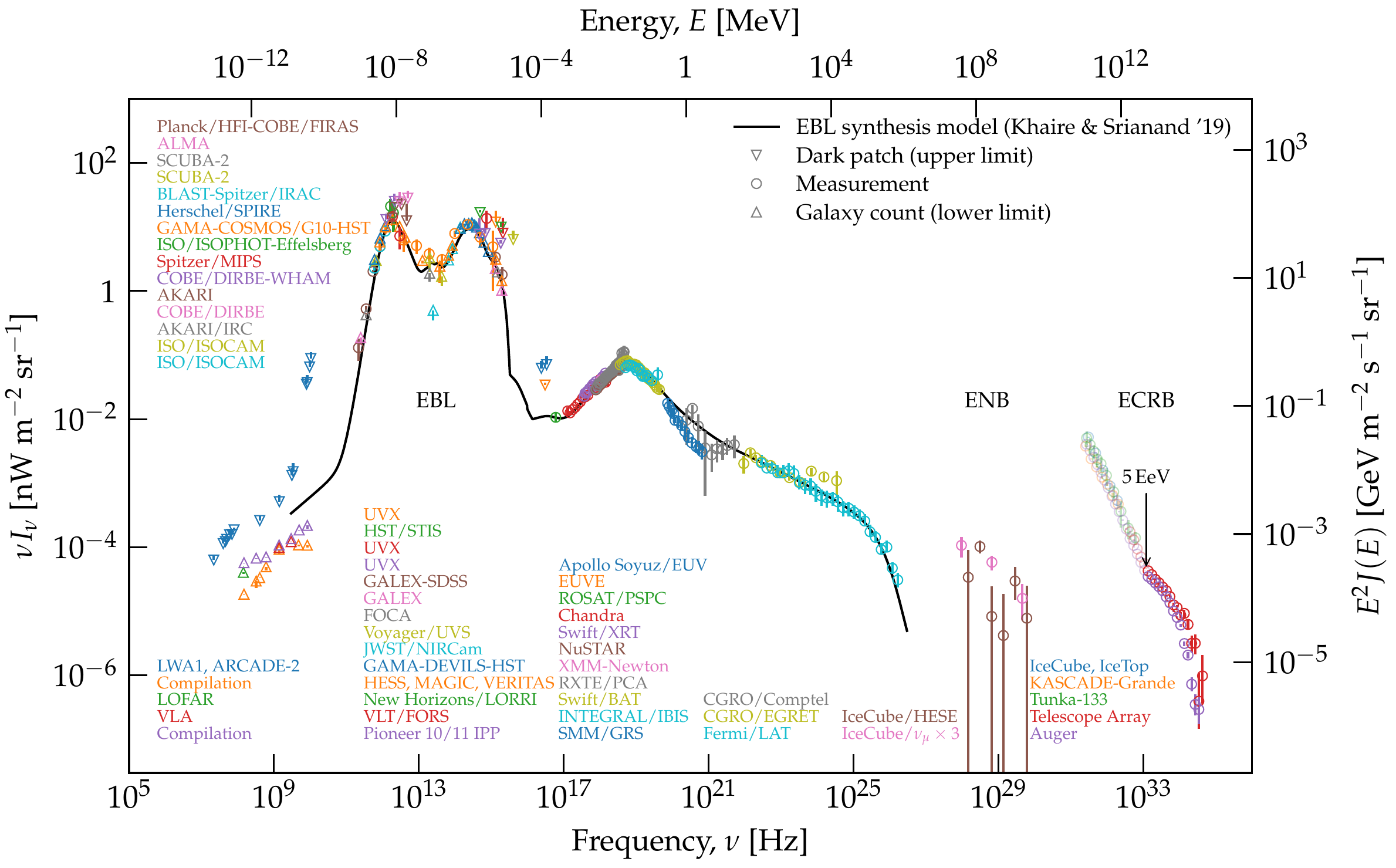}
\caption[]{Multi-messenger spectrum of the universe. Measurements of, or limits on, cumulative emissions from astrophysical sources are indicated by coloured markers, together with the colour-coded names of the corresponding observatories as shown in the figure. \textit{Extracted and adapted from Ref.}~\cite{biteau_2023_7842239}.}
\label{fig:MM_backgd}
\vspace{-0.35cm}
\end{figure}

The astrophysical sources that populate the celestial sphere form a cosmic background through their cumulative emissions. Our understanding of this background has seen tremendous advances since the beginning of this century. The advances are the result of increasingly precise observations over more than 25 energy decades. A better understanding of local foregrounds (the solar system and the Milky Way), exhaustive deep-field counts of galaxies contributing to the cosmic background, and, in bands where possible, absolute measurements of the brightness of dark-sky patches have unveiled the multi-messenger spectrum of the Universe. The measurements of its intensity, in units of power per unit of solid angle on the sky and per unit of detection area, are shown as a function of frequency and energy in Fig.~\ref{fig:MM_backgd}.

Three components corresponding to different messengers are visible in Fig.~\ref{fig:MM_backgd}: the extragalactic background light (EBL), which extends from radio frequencies above 20\,MHz to gamma-ray energies below 1\,TeV, the extragalactic neutrino background (ENB), which has been measured from 40\,TeV to 2.5\,PeV, and the extragalactic cosmic-ray background (ECRB), which reaches energies close to 200\,EeV.\footnote{$1\,\mathrm{TeV} \equiv 10^{12}\,\mathrm{eV}$, $1\,\mathrm{PeV} \equiv 10^{15}\,\mathrm{eV}$, $1\,\mathrm{EeV} \equiv 10^{18}\,\mathrm{eV}$.} Although the underlying question has been the subject of century-long theoretical debates, popularized by Olbers' paradox, the measurement of cosmic backgrounds of astrophysical origin is a young field of research, involving virtually all the astronomical observatories of this century. Particularly noteworthy in the optical band is the recent convergence of direct measurements (New Horizons probe), indirect measurements (H.E.S.S., MAGIC and VERITAS) and galaxy counts (Hubble Space Telescope);\cite{2024ApJ...975L..18G} the measurement of the gamma-ray background up to flux suppression at TeV energies by the \textit{Fermi}-LAT satellite launched in 2008;\cite{2015ApJ...799...86A} the discovery of the ENB in 2013 by the IceCube experiment in Antarctica (see  Ref.~\cite{2024PhRvD.110b2001A} for the latest measurements); and the measurement of the cosmic-ray flux at the highest energies, as accumulated over nearly 20 years at the Pierre Auger Observatory.\cite{2020PhRvL.125l1106A}

The census of astrophysical sources contributing to the EBL, ranging from star-forming galaxies for the optical and infrared components to active galactic nuclei with and without jets for the gamma-ray and X-ray components, has reached a high level of completeness. More than 80\% of the backgrounds can be assigned to known source populations in the eV, keV, and GeV bands. These emissions are driven by star formation, accretion onto supermassive black holes, and ejection around some of these black holes. This knowledge has led to the recent emergence of broadband population synthesis models that reproduce with relative success the full range of EBL observations from the near millimetre to TeV energies (see the black line from Ref.~\cite{2019MNRAS.484.4174K} in Fig.~\ref{fig:MM_backgd}). Comparable knowledge is still lacking for ENB and ECRB due to the low flux of their sources, which goes as $I_\nu$ or $EJ(E)$ i.e., one power of frequency/energy less than in Fig.~\ref{fig:MM_backgd}, and to the angular spread of the charged particles of the ECRB in the Galactic and intergalactic magnetic fields. 

The aim of these proceedings is to give an overview of what is known about the nature and origin of ECRB particles above 5\,EeV, the energy which marks the ankle of the cosmic-ray spectrum, as shown by the arrow in Fig.~\ref{fig:MM_backgd}.\footnote{The cosmic-ray spectrum is shown in Fig.~\ref{fig:MM_backgd} with transparency between the iron knee, around 100\,PeV, and the ankle. Below the iron knee, cosmic rays originate from the Milky Way.} Above the ankle energy, the origin of cosmic rays was already expected to lie outside the Galaxy in the beginning of this century, as illustrated by the proceedings of M. Hillas published in Ref.~\cite{2006astro.ph..7109H}. The author points out that, at the time, there was little convincing evidence for ECRB anisotropies and that the fluxes reported by the various experiments did not agree very well. He therefore gave the newly built Pierre Auger Observatory the task of clarifying the flux level beyond the ankle and of searching for the presence of sources at the highest energies. To quote him: ``such particles cannot have travelled more than about 50\,Mpc, so the magnetic deflections should not be large and there should be few candidate sources in the directional error boxes''. Let us take stock of twenty years of observations led by the Pierre Auger Collaboration.

\section{Cosmic-ray observations at energies larger than 5\,EeV}

The Pierre Auger Observatory is located at 1400\,m above sea level over a vast, flat area of the Argentinean pampas, covering 3000\,km$^2$. The first scientific data collection began in 2004, and the first phase of observations with a complete array ran from 2008 to 2021. Phase 2 is starting at the time of writing with upgraded detectors and will last until at least 2035. This second phase of the Observatory is not covered here. 

Two complementary types of detectors are used to estimate the energy, mass and arrival direction of cosmic rays beyond the ankle. The first one, operating on clear and dark nights with a duty cycle of the order of 15\,\%, allows the Collaboration to reconstruct the longitudinal profile of the energy deposition of cosmic-ray showers in the atmosphere, $\dd E/\dd X$, from the fluorescence of excited dinitrogen molecules. The integral of this profile, measured by the telescopes of the fluorescence detector (FD), provides a calorimetric estimate of the energy, $E$, of the primary cosmic ray. The height of the maximum energy deposition, the slant depth $X_\mathrm{max}$ measured in g\,cm$^{-2}$ as the product of the atmospheric density and traversed length, depends on the nuclear mass $A$ of the cosmic ray via a linear dependence on $\ln (E/A)$, with coefficients determined by hadronic interaction models.\footnote{The particles under consideration, with energies in the EeV range, interact on a fixed target of atoms at rest, with mass energies in the GeV range, i.e., ${\sim}\,(30\,\mathrm{TeV})^2$ in the centre-of-mass frame. This is about a factor of two above the centre-of-mass energy of the Large Hadron Collider.} Although, for the same nature and energy, shower-to-shower fluctuations and measurement uncertainties have so far prevented the assignment of a mass to each primary particle, the first and second $X_\mathrm{max}$ moments (mean and RMS) allow the Collaboration to constrain the mass distribution as a function of energy. The various estimates of the $X_\mathrm{max}$ moments are in good agreement, from the most robust measurements made with the FDs of the Pierre Auger Observatory and of the smaller Telescope Array in the northern hemisphere, to more recent estimates using the impulsive radio-wave signals of the showers or the time traces from particles sampled on the ground (see Ref.~\cite{2023APh...14902819C} for a comparison). Cosmic rays beyond the ankle are atomic nuclei, with a small fraction of ionised hydrogen ($10-15\%$ depending on the hadronic interaction model), which is the tail of the proton distribution at energies just below the ankle, a dominant fraction of helium up to ${\sim}\,20\,$EeV, followed by a dominant fraction of nuclei between carbon and oxygen up to ${\sim}\,50\,$EeV. Beyond this energy, where the flux is suppressed, the number of events detected by the FD becomes too small for mass-distribution inference.

The lack of flux sensitivity of the FD at the highest energies, due to its low duty cycle, is compensated for by the array of water-Cherenkov detectors beneath the volume of atmosphere observed by the FD. The particles in the showers are detected on the ground, day and night, from the Cherenkov emission they induce in cylindrical containers of ultra-pure water. The 1600 water-Cherenkov detectors of the Surface Detector (SD) provide the lateral profile of the showers with a duty cycle close to 100\,\%. The number of particles in the shower depends on the energy of the primary cosmic ray. The interpolated signal at a characteristic distance, namely 1\,km for the water-Cherenkov detectors located on a triangular grid with a spacing of 1.5\,km, is an estimator that is minimally affected by the uncertainties on the lateral profile. This estimator is calibrated to the FD calorimetric energy measurement for jointly detected showers, giving an energy resolution better than 15\,\% beyond the ankle, comparable to the systematic uncertainty on the absolute energy scale of 14\,\%. The arrival times at the water-Cherenkov detectors hit by each shower provide the arrival direction of the primary cosmic ray with an angular resolution better than one degree. Finally, the shape of the Cherenkov signal as a function of time in each detector (rise time, muon-induced spikes) provides composition estimates with a duty cycle close to 100\,\%, albeit at the cost of systematic uncertainties higher than those of FD.\cite{2024arXiv240606315T}

\begin{figure}[t!]
\centering
\begin{minipage}{0.57\linewidth}
\centerline{\includegraphics[width=\linewidth]{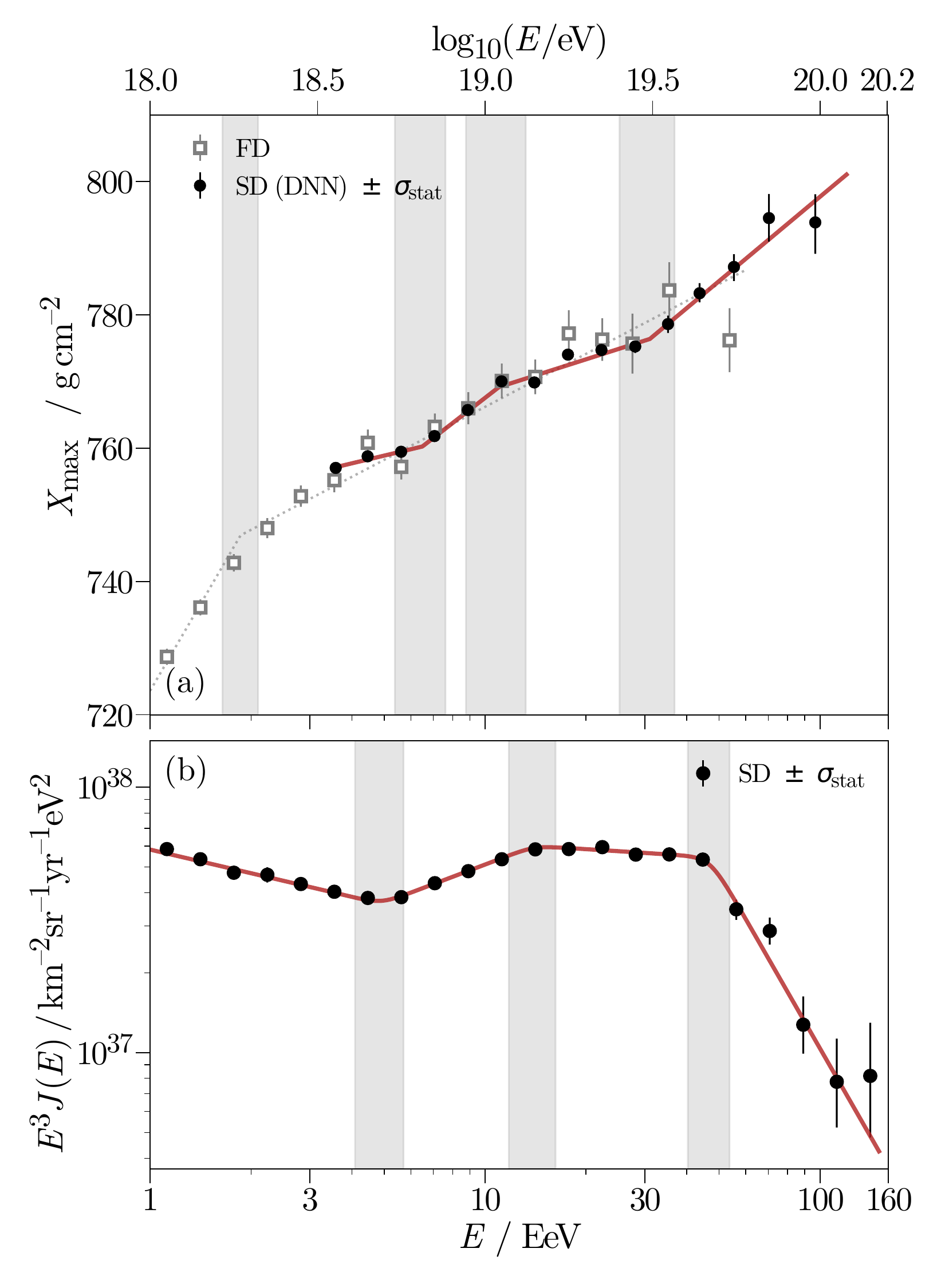}}
\end{minipage}
\hfill
\begin{minipage}{0.39\linewidth}
\centerline{\includegraphics[width=\linewidth]{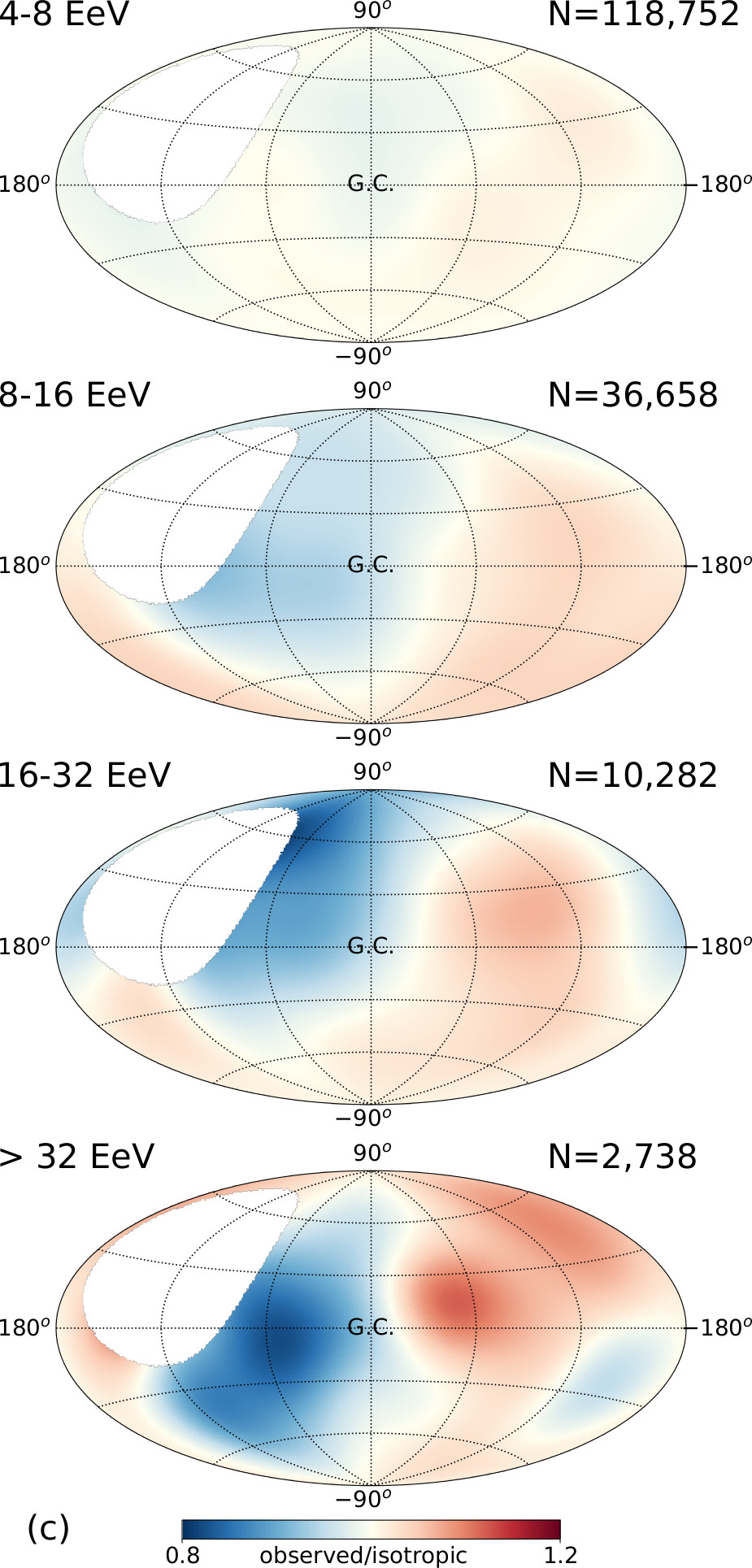}}
\end{minipage}
\caption[]{ECRB observables as a function of energy. Panels (a) and (b) show the mean slant depth and the differential flux multiplied by the third power of energy. The grey bands indicate the position and uncertainties (statistical and systematic) on the break energies observed in the SD data, either for (b) the spectrum or for (a) the mean slant depth from deep neural network (DNN) analysis. Panel (c) shows the cosmic-ray flux normalized to the monopole level in Galactic coordinates in four successive energy bands, with boundaries and observed number events in each band labelled in the figure. The normalized flux is smoothed on an angular scale corresponding to a 45$^\circ$-radius top-hat distribution. \textit{Extracted  and adapted from Ref.}~\cite{2024arXiv240606315T,2024ApJ...976...48A}.}
\label{fig:UHECR_data}
\vspace{-0.35cm}
\end{figure}

The scientific cases covered by SD and FD measurements range from observations of thunderstorms generating electric phenomena in the atmosphere (terrestrial gamma-ray flashes and elves) to upper limits on the flux of neutral cosmic rays (gamma-rays and neutrinos), which constrain the density and decay rate of particles beyond the standard model (e.g., Ref.~\cite{2022arXiv220915310C} for a recent review of the Collaboration's results). We focus here on the key measurements made with the Observatory: the spectrum, composition and arrival directions of cosmic rays beyond the ankle, as shown in Fig.~\ref{fig:UHECR_data}.

The first task entrusted to the Observatory by Hillas was to clarify the level of the cosmic-ray flux and its evolution with energy. Panel (b) of Fig.~\ref{fig:UHECR_data} shows the intensity of the ECRB multiplied by the energy of the cosmic rays, $E^2J(E)\times E$, to better illustrate the observed features. Three slope breaks are observed: the ankle, the instep (recently evidenced in Ref.~\cite{2020PhRvL.125l1106A}) and the flux suppression, which is also called the toe to extend the leg metaphor. The statistical uncertainty on the flux is of the order of one percent between 5 and 15\,EeV, less than 5\,\% up to 40\,EeV and less than 30\,\% below 100\,EeV. Beyond the ankle, the Collaboration has recently identified slope breaks in the mean slant depth at energies close to the spectral slope breaks, as shown in panel (a). The parametrization with three slope breaks above the ankle is preferred to a simple linear model at more than $4\,\sigma$ and is consistent with the synthesis models discussed in the next section. 

Panel (c) of Fig.~\ref{fig:UHECR_data} shows how the sky covered by the Pierre Auger Observatory changes with energy. The arrival directions measured at the Pierre Auger Observatory show an anisotropy level that increases with energy, as illustrated by the increasing contrast from top to bottom. The interplay between the decreasing number of events and the increasing signal amplitude with increasing energy leads to a maximum significance of the deviation from isotropy between 8 and 16\,EeV (dipole amplitude of $6.5 \pm 1.0\,\%$), which was discovered in 2017 and now reaches almost $7\,\sigma$ above 8\,EeV. The map above 32\,EeV in panel~(c)  shows the emergence of structures at smaller angular scales. Searching for excesses without \textit{a priori} knowledge of the angular scale or preferred directions does not yield a significant signal due to the large number of trial factors. However, fixing the centre of the region of interest to the long-suspected Centaurus region, at Galactic coordinates $(l,b) \approx (310^\circ, 20^\circ)$, leads to a flux excess above 40\,EeV and on a $25^\circ$ top-hat angular scale that is significant at $4\,\sigma$.\cite{2022ApJ...935..170A} The second task assigned by Hillas, namely identifying the few sources contributing to the flux in the regime with the lowest magnetic deflections, is not entirely fulfilled at this stage. To do this, the Collaboration uses the knowledge gained on the propagation of cosmic rays in the intergalactic medium and on the populations of sources of the EBL.

\section{Deductions and inferences from cosmic-ray observations}

Unlike the EBL and ENB, which include photon and neutrino emission from stars and active galactic nuclei back to the epoch of reionisation, the ECRB includes emission from such sources only up to a limited horizon. The process limiting the propagation of nuclei beyond the ankle is the interaction with the photon fields populating the universe, mainly the cosmic microwave background and the infrared component of the EBL, on which nuclei photodissociate by emitting one or more nucleons. The magnetic fields traversed induce an angular and temporal spreading of the emitted flux. However, the induced delay does not make a substantial contribution to the total travel time of nuclei in these backgrounds, unless one requires the existence of fields in cosmic voids exceeding ${\sim}\,10\,$nG, which is in tension with cosmological observations. 

\begin{figure}[htb!]
\centering
\includegraphics[width=\linewidth]{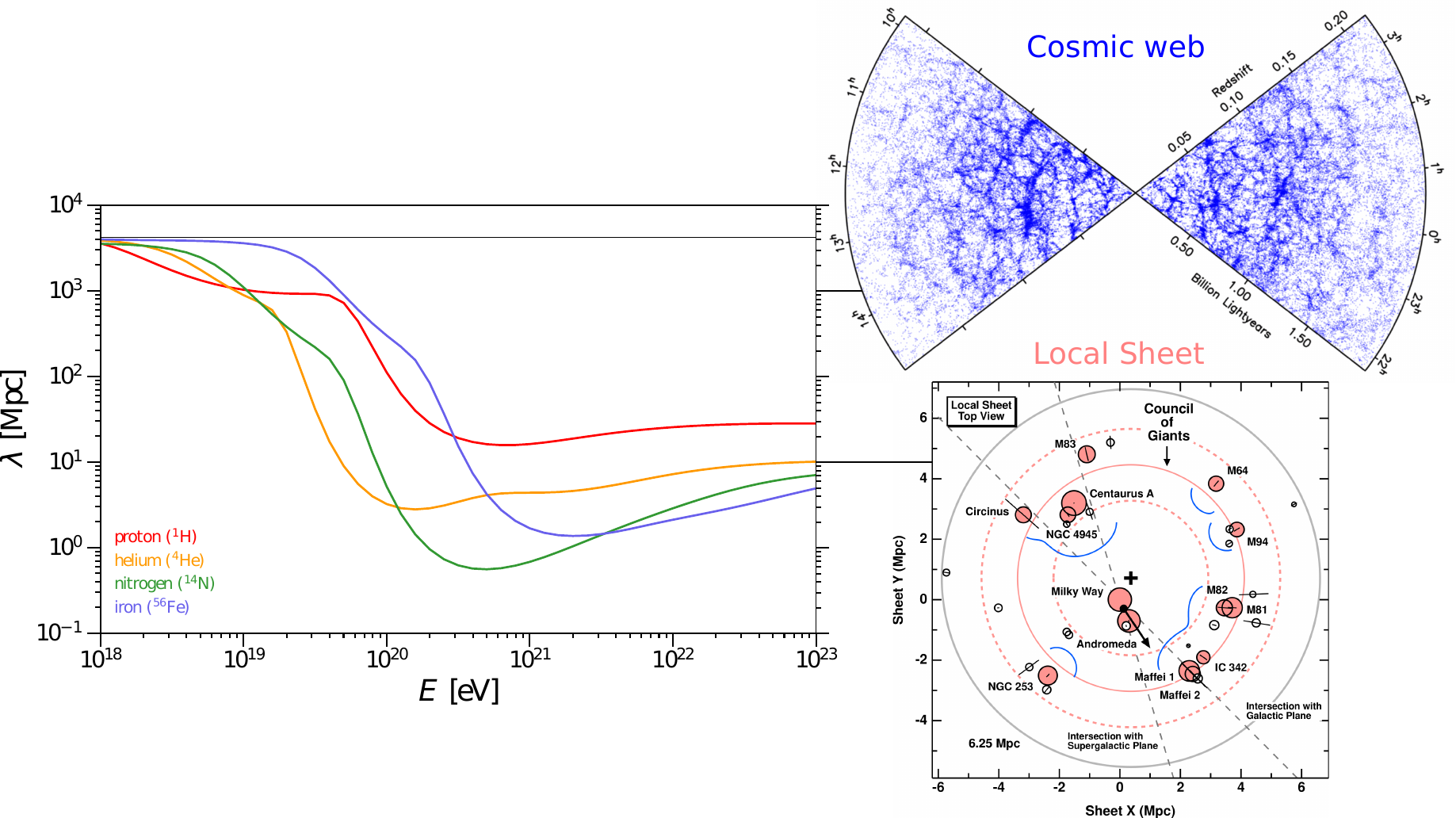}
\caption[]{Energy loss length of nuclei of hydrogen, helium, nitrogen and iron as a function of energy. The horizontal lines associated to the side figures illustrate the distances relevant to large structures of the cosmic web and to the structure in which we reside, the Local Sheet. \textit{Extracted and adapted from Ref.}~\cite{2022PrPNP.12503948A,2005MNRAS.362..505C,2014MNRAS.440..405M}.}
\label{fig:MFP}
\vspace{-0.35cm}
\end{figure}

The energy-loss length of cosmic rays as a function energy is shown for various nuclei in Fig.~\ref{fig:MFP}. The programs that encode the interaction cross-sections of the nuclei and model their cascades in the intergalactic medium have converged over the last decade, so that the uncertainties associated with the propagation are now sub-dominant compared to, for example, the systematic uncertainties in the ECRB measurements. The reduction of the cosmic-ray horizon with increasing energy can be used to qualitatively explain the increase in amplitude of the anisotropies shown in Fig.~\ref{fig:UHECR_data}c. The residual hydrogen nuclei and the helium nuclei around 10\,EeV originate from astrophysical sources located at luminosity distances smaller than 1\,Gpc (redshift $z \approx 0.2$), i.e., from a relatively homogeneous and isotropic portion of the Universe, where a dipolar anisotropy of the order of 10\,\% is observed in the distribution of baryonic mass at near-infrared wavelengths.\cite{2012MNRAS.427.1994G} At higher energies, around 50\,EeV, residual helium nuclei and carbon-to-oxygen nuclei originate from a volume limited to 10\,Mpc and 100\,Mpc, respectively. The latter corresponds to the characteristic size of our supercluster, Laniakea, while the former encompasses the smaller substructure of the cosmic web in which we live: the Local Sheet. The brightest galaxies in the Local Sheet visible from the southern hemisphere lie in a group of galaxies located in the Centaurus region.

To jointly explain the spectrum of the ECRB beyond the ankle, its composition and the evidence for anisotropy from the Centaurus region, synthesis models need to account for foreground galaxies up to about a hundred Mpc, the evolution of the diffuse background from galaxies up to a redshift of $z \approx 1$, the energy losses of cosmic rays during their propagation, and the angular (temporal) dispersion induced by the magnetic fields through which they pass. Models of the Galactic magnetic field are still too divergent to be included in such analyses, although progress is being made in this direction. The recent analysis by the Collaboration,\cite{2024JCAP...01..022A} which combines spectrum, composition and arrival directions beyond 16\,EeV, uses an effective strategy for modelling the magnetic-field effect, parameterized by a smoothing angular scale inversely dependent on the magnetic rigidity $E/Ze$, where $Z$ is the nuclear charge. This analysis confirms the studies carried out by the Collaboration and the community over the past decade: the ECRB beyond the ankle is explained by nuclei escaping from the sources with a particularly narrow rigidity spectrum. This high spectral hardness (in astroparticle jargon) is inferred from the measurement of the slant-depth RMS, which suggests a quasi mono-elemental sequence of nuclei with mass increasing with energy. The conclusion about spectral hardness is independent of the background evolution and of the population of galaxies mapped in the foreground. 

The arrival directions of nuclei with the highest magnetic rigidities can be used to constrain the contribution of foreground galaxies, assuming that coherent deflections by the Galactic magnetic field do not completely erase their flux pattern on the sphere. The strongest correlation comes from the comparison with a catalogue of less than fifty star-forming galaxies at distances up to ${\sim}\,100\,$Mpc. Their 20\,\% contribution to the ECRB flux at 40\,EeV leads to the rejection of isotropy at $4.5\,\sigma$ in favour of excesses with a Gaussian extent of ${\sim}\,20^\circ \times \left(\frac{E/Ze}{10\,\mathrm{EV}}\right)^{-1}$.\cite{2024JCAP...01..022A} This evidence of signal is consistent with the $3.8\,\sigma$ correlation obtained by analysing the arrival directions from the Pierre Auger Observatory alone, and with the $4.5\,\sigma$ correlation when the arrival directions from the smaller Telescope Array are added. Alternatively, the Collaboration has fitted in Ref.~\cite{2024JCAP...01..022A} a foreground consisting of the jetted active galactic nuclei observed in gamma rays by \textit{Fermi}-LAT. Although a contribution of ${\sim}\,15\,\%$ is favoured by the spectral data, it leads to a worse description of the arrival-direction data than an isotropic model, due to the large weight of the nearest blazars.

What do the best-fit synthesis models of the ECRB suggest? First, the hardness of the energy spectra inferred for the nuclei as they escape the sources is very different from the softer spectra expected in the classical model of diffusive shock acceleration. The microphysical origin of this spectral hardness, whether due to an alternative acceleration mechanism and/or to the escape from the magnetised zone where the acceleration and diffusion take place, remains a puzzle, if not a crisis, that has yet to be addressed by the theory and simulations. The upside of this hardness is a significant reduction in the energy budget required for cosmic accelerators. With a soft spectrum, much of the cosmic-ray power emitted by the sources would be at low energies, below the ankle. The hardness of the spectra inferred today, including both protons just below the ankle and nuclei above the ankle, requires an emissivity 25 times less than that inferred by Hillas, who assumed in Ref.~\cite{2006astro.ph..7109H} an ECRB composed entirely of protons with a soft index, $J(E) \propto E^{-2. 4}$.

\begin{figure}[t!]
\centering
\includegraphics[width=0.7\linewidth]{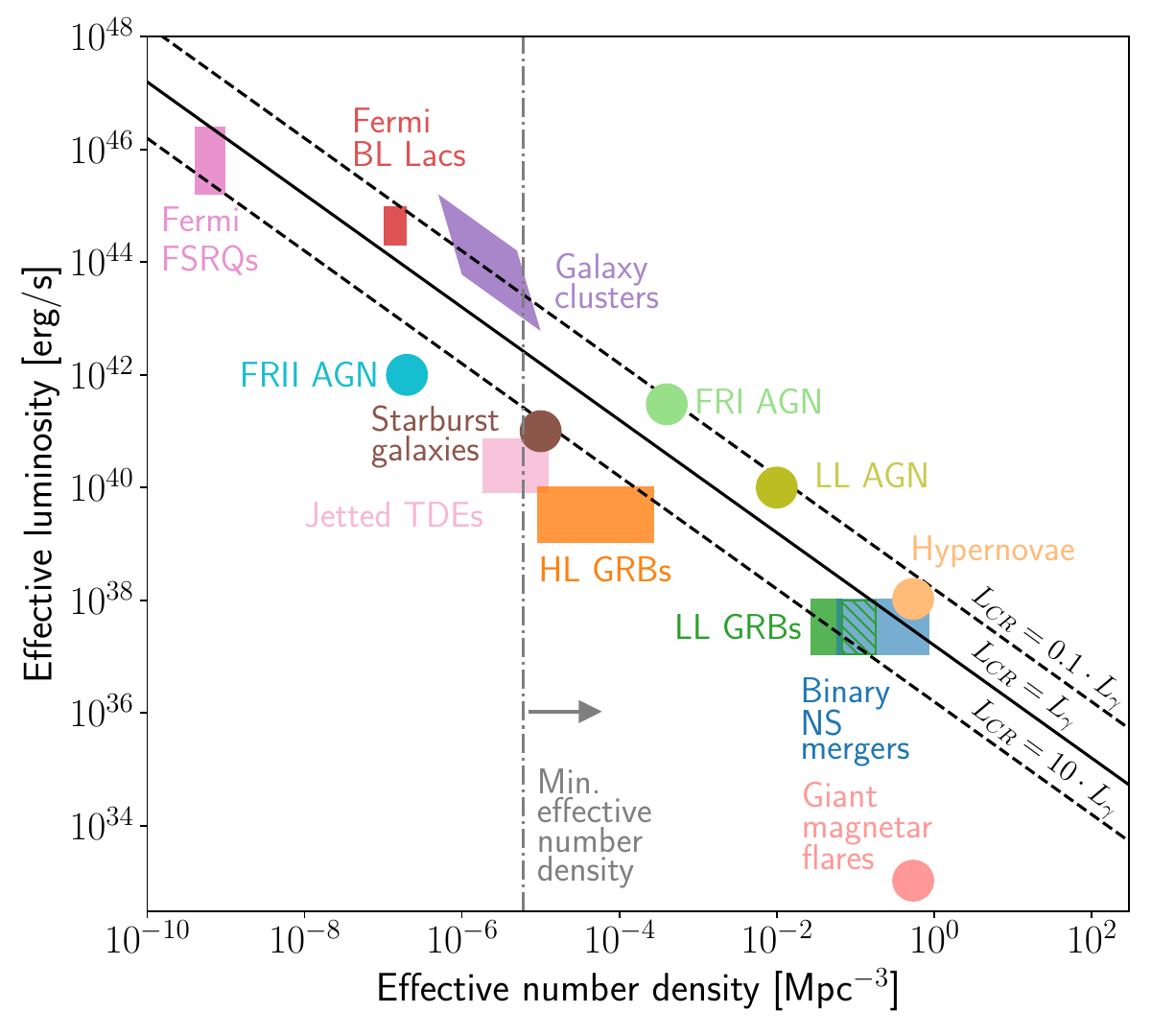}
\caption[]{Effective luminosity as a function of effective number density of candidate populations of sources of the ECRB. The minimum number density not to exceed the level of anisotropy observed in the ECRB is represented by a dotted grey line. \textit{Extracted from Ref.}~\cite{2019FrASS...6...23B}.}
\label{fig:Hillas_eff}
\vspace{-0.35cm}
\end{figure}

Second, pending a correlation at $5\,\sigma$, the evidence for correlation on an intermediate angular scale with star-forming galaxies is in tension with the prior expectation of acceleration of protons and helium nuclei (i.e., $98-99\,\%$ of the baryons in the Universe) by jetted active galactic nuclei. The fact that jetted active galactic nuclei observed by \textit{Fermi}-LAT are not favoured by the arrival-direction data is not in itself surprising. Since the density of their brightest representatives (the BL Lacs and the FSRQ blazars) is low, anisotropies of amplitude much greater than those observed would be expected if these sources were powering the ECRB at the highest energies (see the dotted-dashed grey line in Fig.~\ref{fig:Hillas_eff}, to be compared with the markers at the top left).

The workaround for jetted active galaxies and the workhorse for star-forming galaxies is the not unlikely possibility that the sources of ECRB are transient, although their flux on Earth is constant because of the magnetic fields that cosmic rays pass through. The authors of Ref.~\cite{2022MNRAS.511..448B} have invoked this possibility to explain the anisotropies at the highest energies by the past activity of the jetted active galaxy Centaurus A, reflected as an echo by the magnetised winds of star-forming galaxies in the Local Sheet. The authors of Ref.~\cite{2024ApJ...972....4M} have alternatively suggested transient emission from stellar-sized sources in proportion to the star formation rate of each galaxy. The absence of a signal from the Local Group and the evidence of signal from nearby, bright star-forming galaxies constrain the rate density and energy of the bursts to those of long-duration gamma-ray bursts. It is not yet possible to say with certainty whether star formation or accretion/ejection onto supermassive black holes is at the origin of the ECRB. The composition of the nuclei ejected by the sources, which is surprisingly low in hydrogen and helium, is likely the key to disentangle the remaining contenders.

\section{Summary and outlook}

We have assessed the current understanding of cosmic rays beyond 5\,EeV against the knowledge and expectations of the cosmic-ray community prior to the Pierre Auger Observatory measurements, as described at the time in a review by Michael Hillas. In doing so, we left out many of the scientific cases covered by the Collaboration, including (not least) the nature and origin of the bulk of cosmic rays below the ankle. 

The progress made by the largest cosmic-ray observatory ever built seems to be living up to Hillas' expectations. The cosmic-ray spectrum is measured with exquisite precision from 1 to 100\,EeV, revealing distinct slope breaks at the ankle around $5\,$EeV, the instep around $15\,$EeV and the toe around $45\,$EeV. The slant-depth estimates show that the ECRB flux beyond the ankle is composed of a sequence of nuclei ranging from helium to (at least) oxygen with mass increasing with energy, contrary to previous expectations of a spectrum dominated by a mass fraction of 75\,\% protons and 25\,\% helium. The ECRB synthesis models, although simpler than those developed for the EBL, reproduce the observed spectrum and composition. The modelling of the composition data reduces the emissivity constraints by more than an order of magnitude compared to the inference made early in the century, at the cost of narrow spectra inferred for the nuclei escaping from the sources. These narrow spectra could be a stumbling block for theory. The arrival directions of cosmic rays are not to be outdone. Between the ankle and the instep, the arrival directions show a dipolar anisotropy reaching $7\,\sigma$, confirming the extragalactic origin of the ECRB. Beyond the instep, in the toe region, the correlation with the distribution of star-forming galaxies has reached $4.5\,\sigma$ at the time of writing. The evidence of excess, attributed to galaxies in the Centaurus region a few Mpc away, may be the key to the identification of ECRB sources.

Overall, the most unexpected point is probably the composition of the ECRB, which is so depleted in protons and helium nuclei. The observatory upgrade that led to the start of Phase 2 promises to improve our understanding of atmospheric showers and the nature of primaries to lift the veil on the sources.

%\setlength{\bibsep}{0pt plus 0.3ex}
%\bibliography{bibliography}

\end{document}